\documentclass[aps,prl,superscriptaddress,twocolumn,longbibliography]{revtex4-1}
\usepackage[utf8]{inputenc}
\usepackage{tikz-cd}
\usepackage{color}
\usepackage{dsfont}
\usepackage{bm}
\usepackage{mathtools}
\usepackage{amsthm}
\usepackage{amsfonts,amssymb,amsmath}
\usepackage[hidelinks]{hyperref}
\usepackage{physics} \usepackage{multirow}
\usepackage{graphicx}

\usepackage[export]{adjustbox}
\usepackage{verbatim}
\usepackage{pdfpages} 
\usepackage{pgffor} 

\makeatletter
\AtBeginDocument{\let\LS@rot\@undefined}
\makeatother
\def\supplementfilename{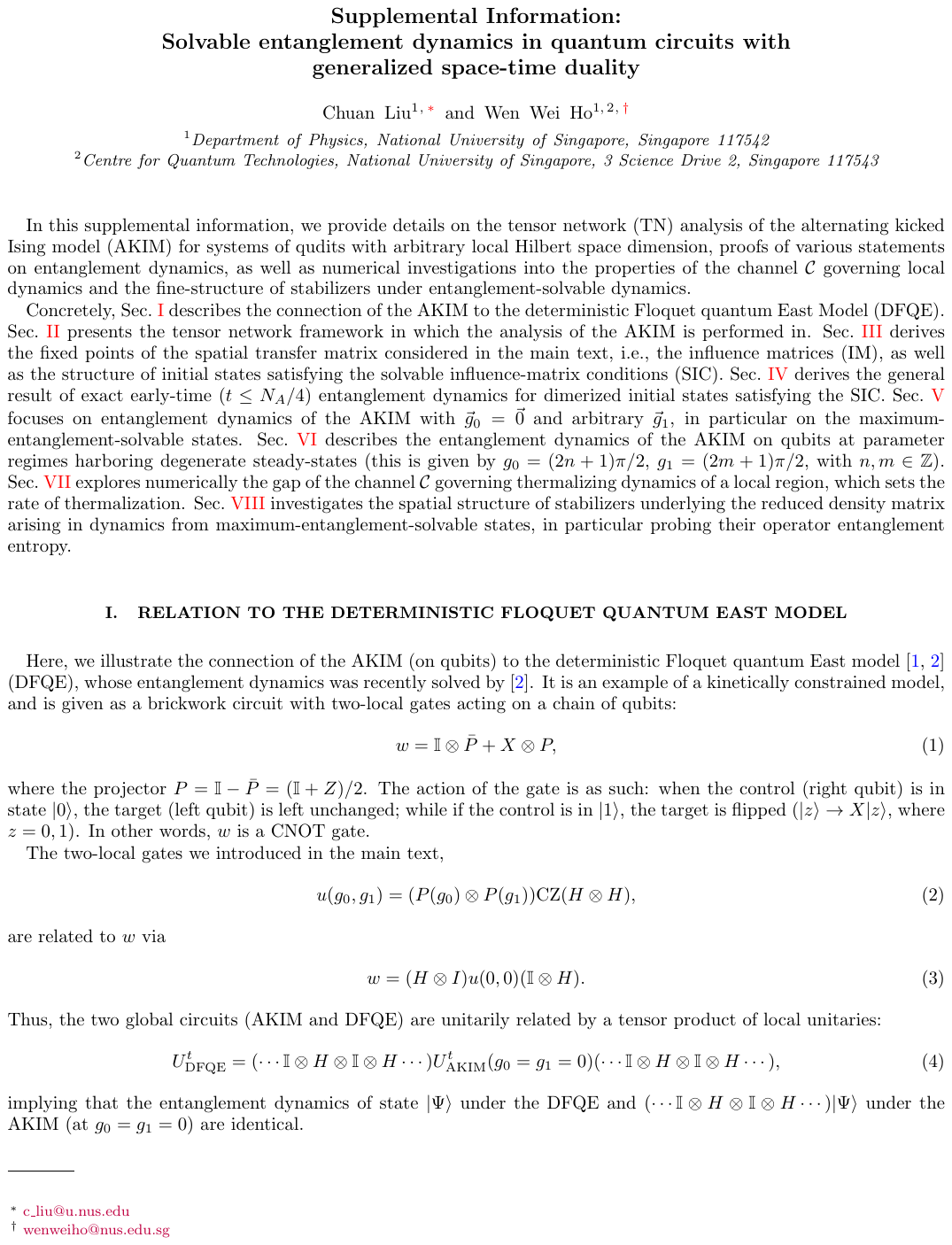}
\pdfximage{\supplementfilename}
\def\numbersupplementpages{\the\pdflastximagepages}

\begin{document}

\title{Solvable entanglement dynamics  
in quantum circuits with \\
 generalized
space-time duality
}

\author{Chuan Liu}\email{c\_liu@u.nus.edu}
\affiliation{Department of Physics, National University of Singapore, Singapore 117551}

\author{Wen Wei Ho}\email{wenweiho@nus.edu.sg}
\affiliation{Department of Physics, National University of Singapore, Singapore 117551}
\affiliation{Centre for Quantum Technologies, National University of Singapore, 3 Science Drive 2, Singapore 117543}

\begin{abstract}
We study the non-equilibrium  dynamics of kicked Ising models in $1$\,$+$\,$1$ dimensions which have interactions alternating between odd and even bonds in time. 
These models can be understood as quantum circuits tiling  space-time with the generalized space-time dual properties of tri-unitarity (three `arrows of time') at the global level, and also second-level dual-unitarity at the local level,  which constrains the behavior of pairs of  local gates underlying the circuit under a space-time rotation.
We identify a broad class of initial product states wherein the effect of the environment on a small  subsystem can be exactly represented by influence matrices with simple Markovian structures, resulting in the subsystem's  full dynamics being efficiently computable.
We further find additional conditions under which the dynamics of entanglement can be solved for all times, yielding rich phenomenology ranging from linear growth at half the maximal speed allowed by locality, followed by saturation to maximum entropy (i.e., thermalization to infinite temperature); to entanglement growth with saturation to extensive but sub-maximal entropy. 
Intriguingly, for certain parameter regimes, we find a nonchaotic class of dynamics which is neither integrable nor Clifford, exemplified by nonzero operator entanglement growth but with a spectral form factor which exhibits large, apparently time-quasiperiodic revivals.
\end{abstract}

\date{\today}
\maketitle


{\it Introduction.}$\,$---$\,$Understanding the dynamics of nonequilibrium quantum many-body systems is one of the central challenges in modern physics: it finds relevance in diverse topics ranging from thermalization~\cite{rigol_thermalization_2008, Nandkishore_2015,dalessio_quantum_2016,  abaninReviewThermalization_2019} to information scrambling~\cite{Hayden_2007,ShenkerBlackHole_2014,   hosur_chaos_2016, roberts_chaos_2017, Mezei2017, LandsmanOTOC_2019}. Dynamics in the far-from-equilibrium regime though is typically difficult to describe, due to the large build-up over time 
of entanglement. Exact classical simulations are limited to small system sizes, while approximate methods like mean-field theory~\cite{MeanReview} and matrix product states~\cite{TEBD, AJDaley_2004, PaeckelTEBD_2019, HemeryMPS_2019} can deal with large sizes  but quickly lose validity at late times. Solvable models of interacting quantum dynamics are thus highly valuable, but rare.  Current well-known examples include integrable systems~\cite{alba_entanglement_2017, EsslerIntegrable_2016, calabrese_entanglement_2020, Klobas54_2021}, Clifford circuits~\cite{NielsenBook, NahumRUC_2017, FarshiClifford_2023, RichterClifford_2023}, as well as 
the recently introduced class of dual-unitary quantum circuits in which local gates are unitary in both space and time directions~\cite{Bertini_2018, 
GopalakrishnanChannel_2019, RatherDU_2020, PiroliDU_2020,ClaeysDU_2021,PhysRevResearch.2.043403}, a realization of a more general concept of space-time duality~\cite{BanulsSpaceTime_2009, HastingsSpaceTime_2015, Bertini_2018, PhysRevX.9.021033, alessioIM_2021, LuSpaceTime_2021, IppolitiSpaceTime_2022} wherein roles of space and time can be interchanged.

\begin{figure}[t]
\includegraphics[width=1\columnwidth]{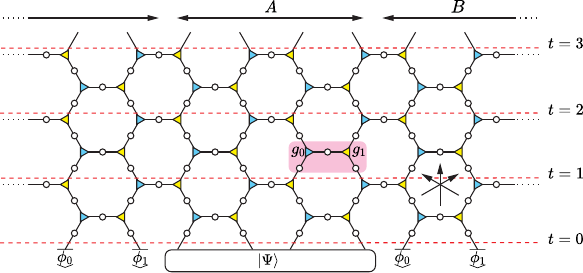}
    \caption{Quantum circuit representation of the AKIM as an (unnormalized) tensor network tiling space-time in a hexagonal fashion, making evident that there are three `arrows of time': `evolution' of the circuit along any one of the black arrows appears unitary. 
    The circuit can be understood as built up from  two-local 2DU gates $u(g_0,g_1)$ (pink shaded box),  where the blue (yellow) triangles on the vertices carry phase factor $g_0 (g_1)$.
    }
    \label{fig:1}
\end{figure}

In this Letter, we introduce a family of  kicked Ising models in 1D that have interactions alternating between odd and even bonds in time, which we call the ``alternating kicked Ising model'' (AKIM).
These models exemplify a general construction yielding quantum circuits with generalized space-time duality beyond dual-unitarity, via tiling of space-time with local gates which endow the system with multiple `arrows of time', tied to the symmetries of the underlying lattice. For the AKIM, this takes the form of a hexagonal lattice, such that it possess the global property of tri-unitarity (three `arrows of time'), but which goes beyond the local tri-unitary framework of Ref.~\cite{Jonay_2021}. We also find that the AKIM possesses the property of `second level dual unitarity'~(2DU), a recent hierarchical generalization of dual-unitarity~\cite{yu2023hierarchical}. 
{Thus, our geometric picture offers complementary insights into constructing models with generalized space-time duality like 2DU and possibly beyond (though see related ideas by Ref.~\cite{PhysRevB.109.214302}).}

We show that because of these properties, the AKIM has local entanglement dynamics that can be  efficiently classically computed, and under certain conditions even analytically solved, displaying rich phenomenology. Interestingly, we find that     dynamics of the AKIM ranges from quantum chaotic, to non-chaotic but which is neither Bethe-integrable nor Clifford. Our results extend our knowledge of solvable interacting quantum dynamics, yielding yet more analytical testbeds to probe interesting physical phenomena like thermalization and beyond \cite{HoExactStateDesign_2022, CotlerStateDesign_2023, IppolitiDeepThermalization2_2022, IppolitiDynamical_2023, shrotriya2023_deep}, quantum chaos \cite{srednicki_chaos_1994, ChanChaos_2018}, and nonequilibrium dynamical phases~\cite{SkinnermIPT_2019, LiMIPT_2019, ChoiMIPT_2020,BaoMIPT_2020, GullansMIPT_2020, ClaeysProjective_2022}.

{\it Model.}$\,$---$\,$Consider discrete-time dynamics  on a 1D chain of qubits, with each  time step $t$\,$\in$\,$\mathbb{N}$ governed by the unitary $
U$\,$=$\,$U_\text{e} U_\text{o}$,
where $U_\text{o/e}$\,$=$\,$e^{-i  H^{(\text{o/e})} } e^{i h \sum_{i=1} \sigma^y_i}$ and 
\begin{align}
H^{(\text{o/e})} =  J \!\!\!\!\!\!\!\!\!\!\!\!   \sum_{ \stackrel{(i,i+1) \in }{ \text{odd/even bonds}} } \!\!\!\!\!\!\!\!\!\!\!\!  Z_i Z_{i+1}  +  \sum_i (g_{i\text{ mod }2}/2+\pi/4) Z_i.
\end{align}
Here $X_i$\,$,$\,$Y_i$\,$,$\,$Z_i$ are Pauli matrices on site $i$. 
 $U$ describes time-evolution alternating between two Ising models which act only on odd or even bonds with strength $J$, subject to dimerized longitudinal fields $g_0,g_1$, and interrupted by global transverse kicks of strength $h$, hence its name ``AKIM''. 
We take $J=h=\pi/4$, and allow $g_0,g_1$ to be arbitrary.
Defined this way, the model is time-periodic i.e.~Floquet. Due to reflection symmetry, dynamics at $(g_0$\,$,$\,$g_1)$ is equivalent to that at $(g_1$\,$,$\,$g_0)$, and so we focus on $0 \leq g_0$\,$\leq$\,$g_1$\,$\leq$\,$2\pi$.

The physical setting of interest is of local thermalizing dynamics following a quantum quench. Concretely, we focus on a small contiguous subsystem $A$ of even number of qubits $N_A$ deep in the bulk and prepared in pure state $|\Psi\rangle$~(see Fig.~\ref{fig:1}); 
and assume the  infinitely-large complement $B$ is prepared  as a dimerized product state $\prod_i |\phi_0\rangle_{2i-1}|\phi_1\rangle_{2i}$.  
We track $A$'s state over time, given by the reduced density matrix~(RDM) $\rho_A(t)$, and desire to understand its entanglement entropy $S(t)$\,$=$\,$-\text{Tr}(\rho_A(t) \log_2 \rho_A(t))$.

{\it Quantum circuit equivalent \& generalized space-time duality.}$\,$---$\,$
Before solving for the AKIM's entanglement dynamics, we note that the AKIM exemplifies a general construction of quantum circuits with generalized space-time duality. To see this, first observe that $U$ is expressible as a brickwork quantum circuit composed of the two-local gates
\begin{align}
    u(g_0,g_1) = \left( P(g_0) \otimes P(g_1) \right) \text{CZ} \left(H \otimes H\right),
\end{align}
where $\text{CZ}$\,$,$\,$H$\,$,$\,$P(g_i)$\,$=$\,$\text{diag}(1$\,$,$\,$e^{ig_i})$ are the standard qubit control-Z,  Hadamard, and phase gates respectively. Note when $g_0$\,$=$\,$g_1$\,$=$\,$0$, $u$ is equivalent (locally) to the CNOT gate, such that the circuit describes the kinetically-constrained deterministic  Floquet quantum East model (DFQE)~\cite{BertiniLocalisedDFQE_2023, bertini2023exact,PhysRevX.10.021051, SI}, whose entanglement dynamics was  recently solved in \cite{bertini2023exact}. Our work advances conceptually that the DFQE lies in a much more general class of entanglement-solvable quantum circuits.

\begin{figure}[t]
    \centering
\includegraphics[width=1\columnwidth,valign = c]{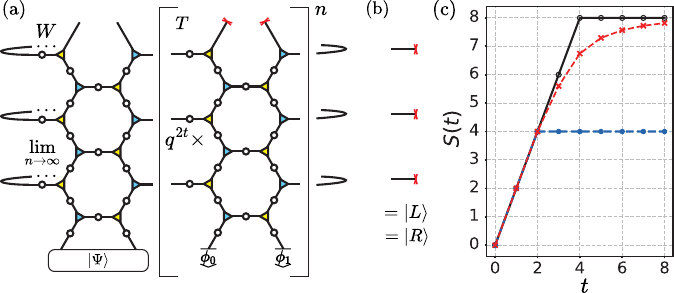}
    \caption{
    (a) Computation of RDM $\rho_A(t)$ using spatial transfer matrices $T$ and temporal-to-spatial map $W$.
    (b) Left and right eigenvectors of $T$: a product of $t$ Bell states.
    (c) Entanglement dynamics of $A$ with $N_A$\,$=$\,$8$. Black line: maximum-entanglement-solvable dimerized product states at $g_0$\,$=$\,$0$ (arbitrary $g_1$), as well as $g_0$\,$=$\,$g_1$\,$=$\,$\pi/2$, showcasing linear ramp with rate $2$ till saturation at maximal entropy.
    Blue dash:  $|z\rangle|z\rangle\cdots$ at $g_0$\,$=$\,$g_1$\,$=$\,$\pi/2$, showing exact linear ramp till saturation at $N_A/2$.  
    Red dash:    $|z\rangle|z\rangle\cdots$ at  $(g_0$\,$,$\,$g_1)$\,$=$\,$(5\pi/16, 7\pi/16)$. Only at early times $t$\,$\leq$\,$N_A/4$ is there an exact linear ramp; beyond which the approach to maximum entropy becomes exponential.}
    \label{fig:2}
\end{figure}

We then introduce a tensor network (TN) representation, using basic tensors 
\begin{equation}
    \includegraphics[height=1cm,valign = c]{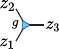} 
    = \delta_{z_1z_2z_3} e^{ig^{(z_{1})}}, 
    \quad 
    \includegraphics[height=1cm,valign = c]{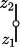} = (-1)^{z_1 z_2}/\sqrt{2},
    \label{eqn:basic_tensors}
\end{equation}
which describe a 3-legged Kronecker delta with phase $g$ and Hadamard respectively ($z_i$\,$\in$\,$\{0,1\}$). Using standard rules of TN manipulations, we can build up the local gate $u(g_0$\,$,$\,$g_1)$ as shown in Fig.~\ref{fig:1}, and hence the global circuit $U^t$ for time $t$~(see~\cite{SI} for details). 

We find that $U^t$ describes a {\it tiling of space-time by a decorated hexagonal lattice}.
The diagrammatic representation makes immediately manifest that there are three ``arrows of time'', due to similarity of the tensor network under $120^\text{o}$ rotations: perpendicular to  any one of the black arrows in Fig.~\ref{fig:1}, the  ``evolution'' of the circuit is unitary. This property can be termed `tri-unitarity'~\cite{Jonay_2021}, though we stress it is present at the global circuit level and not at the local gate level (there is no obvious `blocking' of local gates to form a tri-unitary gate). 
We  instead observe the local gates possess a recently-proposed relation governing  contractions of pairs of them along the space direction, termed ``second-level dual unitarity''(2DU)~\cite{yu2023hierarchical}:  
\begin{equation}
    q\includegraphics[height=1.5cm,valign = c]{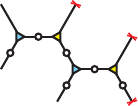} =  \includegraphics[height=1.5cm,valign = c]{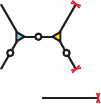}, q\includegraphics[height=1.5cm,valign = c]{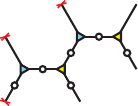} = \includegraphics[height=1.5cm,valign = c]{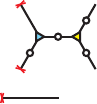},
    \label{eqn:2DU}
\end{equation}
which endows the system with analytic tractability in the computation of spatiotemporal correlations. 
Above, we have introduced bold tensors representing the folding of forward and backward branches of evolution, such as
\begin{equation}
    \includegraphics[height=1cm,valign = c]{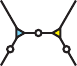} = \includegraphics[height=1cm,valign = c]{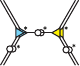},
     \qquad 
    \includegraphics[height=0.4cm]{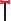} = \includegraphics[height=0.4cm]{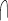},
\end{equation}
which are the folded local gate $u\otimes u^*$ and  trace operation respectively.

Importantly, while the discussion for the AKIM  described a hexagonal lattice in (1+1)D,  the idea behind its construction is quite general: for example, if instead we considered tiling of  space-time with gates of Eq.~\eqref{eqn:basic_tensors}  using a square lattice, then this describes a model with (standard) dual-unitary (two arrows of time, from  the 90$^\text{o}$ rotation symmetry): the canonical self-dual kicked Ising model~\cite{Bertini_2018, HoExactStateDesign_2022}. However, one may repeat the same construction in higher spatial dimensions (see \cite{PhysRevB.109.214302} too); further, the tiling of space-time need not be regular in flat space---for example, it could be a hyperbolic lattice~(see e.g.~\cite{Koll_r_2019,PhysRevLett.128.013601}).
The construction extends too to systems of qudits through the natural qudit generalization of the phase and Hadamard gates, see the Supplemental Material~\cite{SI}.
Our proposal thus represents an appealing complementary geometrical picture to generate quantum circuits with generalized space-time duality like 2DU and possibly beyond.

{\it Exact influence matrices.}---We now return to the AKIM and demonstrate how its generalized space-time dual properties allow us to understand not only dynamics of few-point correlation functions (as was shown already in \cite{Jonay_2021} and \cite{yu2023hierarchical}), but also solve for its entanglement dynamics, which is much more complex. {We note that because of its triunitary and 2DU properties, many aspects of its dynamics follow from \cite{Jonay_2021, yu2023hierarchical}, but because of the model's underlying geometric construction, it possesses additional structure allowing us to fully solve its dynamics under some conditions, as we shall see.}

First, we note the RDM $\rho_A(t)$ can be computed by raising to a high power `spatial transfer matrices' $T$, inserting the operator $W$ mapping temporal information to the spatial region $A$, before tracing over the temporal degrees of freedom, see Fig.~\ref{fig:2}(a). 
Due to unitarity and locality of the original dynamics, it can be shown that $T$ has only a single eigenvalue $1$, with all other eigenvalues $0$ with Jordan blocks bounded in size by $2t$, so that $T^{n \geq 2t} = |R\rangle \langle L|$, where $\langle L|, |R\rangle$ are  the left and right eigenvectors respectively satisfying $\langle  L|T=\langle L|$ and $T|R\rangle=|R\rangle$ \cite{alessioIM_2021,bertini2023exact}. 
These eigenvectors are also called ``influence matrices''(IM), as they effectively encode the effect of the bath on the subsystem as a quantum state living on the temporal direction \cite{alessioIM_2021, SonnerIM_2021,GiacomoIM_2022}.
The computation of $\rho_A(t)$ in a thermodynamically large system thus simplifies to 
$\langle L|W|R\rangle$. 

Generally, the form of an IM  in a strongly-interacting system is complicated and has no closed form expression \cite{AlessioBarrier2_2021,SonnerIM_2021, FolignoIM_2023}. However, for the AKIM, if we impose that the dimerized product state $|\phi_0\rangle |\phi_1\rangle \cdots$ on $B$ satisfies the {\it solvable IM conditions}~(SIC): 
\begin{equation}   \includegraphics[height=1.35cm,valign = c]{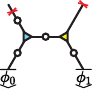} =\frac{1}{q^2} \includegraphics[height=0.5cm,valign = c]{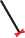},\qquad \includegraphics[height=1.35cm,valign = c]{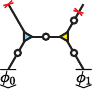} =\frac{1}{q^2} \includegraphics[height=0.5cm,valign = c]{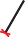},
\label{eqn:SIC}
\end{equation}
we find   $|L\rangle$\,$=$\,$|R\rangle$ and admits the simple form of a product of $t$ Bell pairs, see Fig.~\ref{fig:2}(b) and~\cite{SI}.
This IM for the AKIM is identical to that of dual-unitary circuits (within its corresponding solvable states~\cite{PiroliDU_2020}): it has zero temporal entanglement, such that the bath's effect is a perfect Markovian dephaser on the boundaries. 
Proving this   involves straightforward TN manipulations~\cite{SI}, for example deriving $T|R\rangle$\,$=$\,$|R\rangle$  for $t$\,$=$\,$2$ proceeds as follows:
\begin{equation}
\includegraphics[height=2.7cm,valign = c]{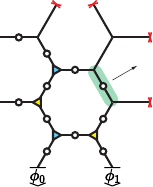} = \includegraphics[height=2.7cm,valign = c]{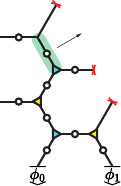} =\includegraphics[height=2.2cm,valign = c]{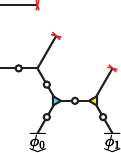},
\end{equation}
where we used unitarity along the black arrows to remove the green-shaded gates, before invoking SIC.

We  thus establish our first result for the AKIM:   the RDM can be computed exactly as $\rho_A(t)$\,$=$\,$\mathcal{C}^t\left[|\Psi\rangle_A\langle \Psi|_A\right]
$ where the quantum channel $\mathcal{C}$ is
\begin{align}
   \mathcal{C} = 2^{N_A-1} \includegraphics[height=1.5cm,valign = c]{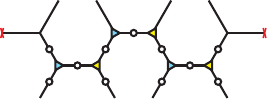}.
    \label{eqn:C_channel}
\end{align}
Note $\mathcal{C}$ is unital: $\mathcal{C}[I_A]$\,$=$\,$I_A$, i.e., the infinite-temperature state $\rho_{T=\infty}$\,$=$\,$I_A/2^{N_A}$ is a steady-state, though there can be more.
Physically, $\mathcal{C}$ describes   evolution under the AKIM on region $A$ for one time-step, followed by dephasing on the boundaries. 
We emphasize this result amounts to a tremendous reduction of the complexity of the problem, as it entails an efficient $O(N^0)$ classical protocol to compute local  dynamics even in a thermodynamically large  system.

{\it Solvable entanglement dynamics}$\,$---$\,$
Next, we consider entanglement entropy generation within $A$. 
We can divide the analysis into two cases: (i)~early times $t$\,$\leq$\,$N/4$, when the TN factorizes into two disjoint diagrams due to causality; and (ii)~late-times $t$\,$>$\,$N/4$.

For early times (i), for {\it any} $(g_0$\,$,$\,$g_1)$,
we find that if states in $A$ also satisfy SIC, then we can prove $S(t)$\,$=$\,$2t$ {which has also been independently obtained in  parallel work~\cite{Foligno2DU_2023}.}
Diagrammatically, this is done by fully contracting the shallow TN corresponding to the generalized purity $\text{tr}(\rho^n_A(t))$ for all $n$\,$\in$\,$\mathbb{N}$ and taking the limit $\lim_{n\to 1}S_n(t)$ where $S_n(t)$ is the $n$-th R\'enyi entropy~\cite{SI}.

For late-times (ii), we can further distinguish the following cases which go beyond the regime discussed in~\cite{Foligno2DU_2023}. 
Case (a): when $g_0$\,$=$\,$0$ or $\pi$ with arbitrary $g_1$,
 we identify a set of local TN identities enabling us to prove $\mathcal{C}^{N_A+1}$\,$=$\,$\mathcal{C}^{N_A}$,  and that $\rho_{T=\infty}$ is the only non-trivial eigenvector~\cite{SI}. 
This implies  {\it any} initial state $|\Psi\rangle$ thermalizes {\it exactly} to infinite-temperature in at most  $t$\,$=$\,$N_A$. 
However, there are states which equilibrate faster: if dimerized product states on $A$ satisfy SIC and additionally {\it solvable entanglement conditions} (SEC):
\begin{equation}
    \includegraphics[height=1.35cm,valign = c]{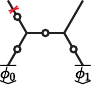} = \includegraphics[height=1.35cm,valign = c]{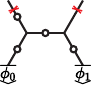}, 
    \label{eqn:SEC}
\end{equation}
where the red cross is  
\begin{equation}
\includegraphics[height=0.75cm,valign = c]{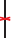} = \includegraphics[height=0.75cm,valign = c]
    {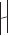} =\sum_{z} \includegraphics[height=1.35cm,valign = c]{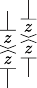}~,
\end{equation}
a projector pinning the forwards and backwards branches into the same computational basis state,  
then provably  $S(t)$\,$=$\,$\min(2t,N_A)$, see Fig.~\ref{fig:2}(c)~\cite{SI}. These ``maximum-entanglement-solvable'' states are exhausted by $|\phi_0\rangle|\phi_1\rangle$\,$\in$\,$\{|z\rangle \otimes e^{i\theta X}|0\rangle\}$\,$\cup$\,$\{|\pm\rangle \otimes e^{i \theta' Z}|+\rangle\}$, where $X|\pm\rangle$\,$=$\,$\pm|\pm\rangle$.
We note the rate of entanglement generation is half the maximum possible in local brickwork circuits, achieved in dual-unitary circuits~\cite{TianciVelocity_2022}. 
This is consistent with the results of recent work exploring quantum information propagation speeds in 2DU circuits~\cite{Foligno2DU_2023}~(here, their $n_\Lambda$\,$=$\,$d$,  predicting an asymptotic
entanglement velocity  $v_E$\,$=$\,$1/2$; however we stress our finding holds even non-asymptotically). 

Case (b): when $g_0$\,$=$\,$g_1$\,$=$\,$\pi/2$,
(or $3\pi/2$), instead another set of local TN identities hold so that $\mathcal{C}^{N_A/2+1}= \mathcal{C}^{N_A/2+2}$. This implies that system equilibrates exactly at finite time $t$\,$=$\,$N_A/2$\,$+$\,$1$ for a generic state.
However, interestingly,  $\rho_{T=\infty}$ is now not the only steady-state --- the equilibrium state need not have maximal entropy. To understand what the additional steady-states are, we show in~\cite{SI} the late-time channel has the TN
\begin{equation}
    \mathcal{C}^{N_A/2+1}  \propto \includegraphics[height=1.5cm,valign = c]{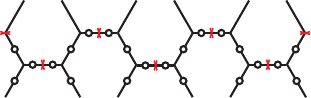}, 
\end{equation}
and read-off  that $\mathcal{C}^{N_A/2+1}=b_1 \circ b_{N_A} \circ
\prod_{\text{even bonds}}
c_{i,i+1}
\circ \prod_{\in \text{odd bonds}} d_{i,i+1}$, where $b_i$ is the local $Z$-dephasing channel, and $c(d)_{i,i+1}$ are the two-site $ZZ(XX)$-dissipative channels. For example, $c_{i,i+1}[\rho]$\,$=$\,$\frac{1}{2} ( \rho$\,$+$\,$Z_i Z_{i+1} \rho Z_i Z_{i+1} )$. It is straightforward to check there is a super-extensive set of $2^{N_A-1}$ independent operators invariant under $\mathcal{C}$, given by all  products of $Z_i Z_{i+1}$ on odd bonds and $X_i X_{i+1}$ on even bonds.
Focusing on the family of  dimerized initial states $|\phi_0\rangle|\phi_1\rangle\in\{|z\rangle \otimes e^{i\theta Z}|+\rangle\}\cup\{|y,\pm\rangle \otimes e^{i \phi X}|0\rangle\}$  and those related by $\phi_0$\,$\leftrightarrow$\,$\phi_1$ (here $Y|y,\pm\rangle$\,$=$\,$\pm|y,\pm\rangle$), we can show that these states have zero expectation values within the conserved quantities and moreover have exact linear growth of entanglement $S(t)$\,$=$\,$\min(2t,N_A)$. We dub these``maximum-entanglement-solvable'' states too. For $|z\rangle|z\rangle\cdots$ instead 
$S(t)$\,$=$\,$\min(2t,N_A/2)$. These are illustrated in Fig.~\ref{fig:2}(c).

Case (c): when  $(g_0,g_1)$\,$=$\,$(\pi/2$\,$,$\,$3\pi/2)$, the situation is richer as we find $\mathcal{C}$ has now some eigenvectors with eigenvalue $-1$, indicating that the system generically  does not equilibrate but rather oscillates at late times! This stems from the facts that (i) states staisfying SIC  at   $(\pi/2$\,$,$\,$\pi/2)$  are identical to those at $(\pi/2$\,$,$\,$3\pi/2)$, and (ii) the action of the channel  in the latter case (called $\mathcal{C}_{1/2,3/2}$)  applied $t$ times  is equal to that of the former (called $\mathcal{C}_{1/2,1/2}$)  applied $t$ times, {\it followed} by the action of a Pauli channel which cycles between $IYIY\cdots$\,$,$\,$YYYY$\,$,$\,$\cdots$\,$,$\,$YIYI\cdots$\,$,$\,$IIII\cdots$ beginning at $t = 1$. 
This arises from the circuits' Clifford nature~\cite{SI}.
Since we know the invariant operators of $\mathcal{C}^t_{1/2,1/2}$ at late times ($t$\,$\geq$\,$N_A/2+1$), which  all  commute with $YYYY\cdots$ but not all with $IYIY\cdots$ or $YIYI\cdots$, it follows  that  $\mathcal{C}_{1/2,3/2}^{t}$\,$=$\,$\mathcal{C}_{1/2}^{t}$ and $\mathcal{P}_{IY}\circ\mathcal{C}_{1/2,1/2}^{t}$ for late even and odd $t$  respectively, explaining the generic oscillatory behavior. 
However, since  a Pauli channel is non-entangling,  entanglement dynamics at $(\pi/2,\pi/2)$ and $(\pi/2,3\pi/2)$ will be identical for the same state.

Lastly, for generic $(g_0$\,$,$\,$g_1)$ away from cases (a-c), we  numerically find  $\rho_{T = \infty}$ is the unique unity eigenvector of $\mathcal{C}$, while  there are eigenvectors with non-zero eigenvalues $\lambda$. The spectral gap   $\Delta$\,$:=$\,$1$\,$-$\,$\max_{\lambda:|\lambda|<1} |\lambda|$  determines the rate of thermalization (see~\cite{SI} for a numerical analysis). We also do not find any obvious simple product states whose entanglement dynamics can be analytically understood beyond the early-time regime~(Fig.~\ref{fig:2}(c)).


\begin{figure}[t]
\includegraphics[width=1\columnwidth]{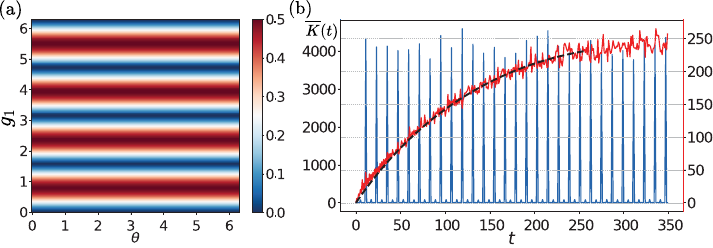}
    \caption{(a)~Operator entanglement of a stabilizer at intermediate times underpinning the RDM $\rho_A(t)$, as a function of phase $g_1$ and angle $\theta$ parameterizing the initial state. 
    The stabilizer generally has non-trivial operator entanglement entropy, indicating complex structure in space that cannot be exhibited by Pauli stabilizers. 
    (b)~SFF of the AKIM at $N_A$\,$=$\,$8$. Blue (left y-axis): SFF for $g_0$\,$=$\,$0$,  shows regular revivals. 
    Red (right y-axis): SFF for generic parameters falls within the COE class: the fit is to the RMT prediction $2t$\,$-$\,$t\ln(1+2t/2^{N_A})$ \cite{MehtaBook}. Note the differences in scale. 
    }
    \label{fig:3}
\end{figure}

{\it Discussion.}--We have introduced the AKIM, and through it demonstrated a general geometrical construction of quantum circuits possessing generalized space-time duality via the idea of space-time tiling. 
This construction recovers existing space-time dualities, like dual-unitarity~\cite{Bertini_2018, 
GopalakrishnanChannel_2019, RatherDU_2020, PiroliDU_2020,ClaeysDU_2021,PhysRevResearch.2.043403} and second level dual-unitarity~\cite{yu2023hierarchical}.
For the AKIM, we further showed how because of its generalized space-time dual properties,  
its entanglement dynamics can be efficiently and at times fully  solved. 

It is interesting to contrast our findings to other known analytically-solvable models of quantum dynamics.
To the best of our knowledge, our model is not generically integrable nor  Clifford~(namely, dynamics which preserve the Pauli group~\cite{NielsenBook, NahumRUC_2017, FarshiClifford_2023, RichterClifford_2023}).
However, our calculation of the generalized purities in the entanglement solvable cases (a) $g_0 = 0$ or $\pi$ with arbitrary $g_1$, (b) $g_0=g_1=\pi/2$ or $3\pi/2$, (c) $g_0=\pi/2$ and $g_1 = 3\pi/2$ yields that the entanglement spectrum of $\rho_A(t)$ is flat~\cite{SI}, a feature exhibited too by Clifford dynamics beginning from stabilizer states~\cite{NahumRUC_2017}. On the other hand, we find that only parameters $(g_0$\,$,$\,$g_1)$\,$=$\,$(n\pi/2$\,$,$\,$m\pi/2)$ of the AKIM where $n$\,$,$\,$m\in\mathbb{Z}$, constitute Clifford circuits, while we have shown that full entanglement-solvability extends beyond these isolated points --- namely, a large parameter range within case (a). Moreover, even when the AKIM is  Clifford, as in case (b,c) and specific points in case (a), the entanglement-solvable states need {\it not} be stabilizer states, a condition otherwise needed in order for the Gottesman-Knill theorem of efficient classical simulability to apply \cite{NielsenBook,Gottesman}. Thus, solvability in these cases does not seem tied to the circuit being Clifford. 

To drive home the difference of the entanglement solvable cases of the AKIM with Clifford evolution, we  investigate the decomposition of the RDM at various times for  case (a): $g_0$\,$=$\,$0$ and arbitrary $g_1$, beginning from dimerized initial states $|0\rangle\otimes e^{i\theta X}|0\rangle$ on $A$,  into the product of  projectors onto orthogonal eigenspaces $\rho_A(t)$\,$\propto$\,$\prod_i (1+O_i)$. Here the stabilizers $O_i$ are defined to be mutually commuting, each having equal numbers of $\pm 1$ eigenvalues, but need not be a Pauli string.  Fig.~\ref{fig:3}(a)~shows a characterization of the spatial structure of a representative $O_i$ at a later time: it generically develops operator entanglement---entropy related to  non-factorizability into a tensor product of locally-supported operators~(details in~\cite{SI}), which can never happen under Clifford evolution~\cite{fattalStabilizerEntanglement_2004, NielsenBook}. Moreover, 
the spectral form factor~(SFF) $\bar{K}(t)$\,$=$\,$|\text{tr}(U^t)|^2$ \cite{Bertini_2018, LiuSFF_2018} for $g_0$\,$=$\,$0$  (averaging over spatially-random $g_1$ uniformly \footnote{We note each instance of the circuit from this ensemble has entanglement dynamics which is still analytically solvable.}), {exhibits surprisingly large, apparently time-quasiperiodic revivals, which conforms neither to}  Poissonian, Wigner-Dyson dynamical classes, nor those of Clifford circuits~(see e.g.~\cite{FarshiClifford_2023}). 
To the best of our knowledge, this represents an interesting new kind of non-chaotic yet non-integrable, non-Clifford quantum dynamics deserving further investigation.
In contrast, uniformly averaging over  $g_0$\,$,$\,$g_1$ yields an SFF agreeing perfectly with the circular orthogonal ensemble~(COE) of random matrix theory~(RMT)~\cite{MehtaBook}.  All these indicate that the physics of the AKIM is very rich.
{It would be interesting to analyze the dynamics of other generalized space-time dual models, generated similarly from the geometric construction we have put forth in this work.}  

\begin{acknowledgments}
{\it Acknowledgments.}~We thank Katja Klobas, Bruno Bertini and Huang Qi for interesting discussions. W.~W.~H. is supported by the Singapore NRF Fellowship, NRF-NRFF15-2023-0008, and the CQT  Bridging Fund. 
\end{acknowledgments}
 
\bibliography{main}

 \foreach \x in {1,...,\numbersupplementpages}
    {
        \clearpage
        \includepdf[pages={\x,{}}]{\supplementfilename}
    }
    
\end{document}